\documentclass[conference]{IEEEtran}
\IEEEoverridecommandlockouts
\usepackage[numbers]{natbib}
\usepackage{amsmath,amssymb,amsfonts}
\usepackage{algorithmic}
\usepackage{graphicx}
\usepackage{textcomp}
\usepackage{xcolor}
\usepackage{hyperref}

\usepackage{floatrow}
\usepackage[caption=false]{subfig}

\usepackage{cuted}

\usepackage{flushend}

\usepackage[]{algorithm2e}
\usepackage{physics}

\def\BibTeX{{\rm B\kern-.05em{\sc i\kern-.025em b}\kern-.08em
    T\kern-.1667em\lower.7ex\hbox{E}\kern-.125emX}}

\begin{document}
\title{Short Text Topic Modeling:\\Application to tweets about Bitcoin}

\author{\IEEEauthorblockN{ Hugo Schnoering}
\IEEEauthorblockA{\textit{Napoleon Group} \\
hugo.schnoering@napoleon-group.com}
}
\maketitle

\begin{abstract}
Understanding the semantic of a collection of texts is a challenging task. Topic models are probabilistic models that aims at extracting "topics" from a corpus of documents. This task is particularly difficult when the corpus is composed of short texts, such as posts on social networks. Following several previous research papers, we explore in this paper a set of collected \textit{tweets} about bitcoin. In this work, we train three topic models and evaluate their output with several scores. We also propose a concrete application of the extracted topics. 
\end{abstract}

\section{Introduction}

Understanding the semantic of a collection of texts, or \textit{corpus}, is a challenging task. Topic modeling provides a convenient way to analyze large collections of unstructured text, and are one of the most popular methods for learning representations of textual data. The increasing amount of data  requires the use of unsupervised algorithms that learn meaningful patterns of words without requiring any prior annotations.  From word co-occurrences, topics models aims at extracting topics from a set of documents, that is, groups of words that tend to appear together. Various hierarchical probabilistic graphical models such as Latent Dirichlet Allocation (LDA) \cite{blei2003latent}, Probabilistic Latent Semantic Allocation (PLSA) \cite{hofmann2013probabilistic},  and hundreds of variants, have been introduced to this end. In PLSA and LDA, topics are distribution vectors over the vocabulary and documents consists of a mixture of topics. \\

Social networks, such as Reddit or Twitter,  have become important communication tools for people, among retail investors. Previous studies show relations between social network data and performance of securities or other financial instruments \cite{2011,  colianni2015algorithmic, mao2013twitter, sul2014trading}. In particular, social networks are an important source of information for cryptocurrency users and investors. Data from social networks could be used to detect breaking news or trending projects. It is however worth pointing out that messages on those platforms are most often characterized by their lack of context and their flexible language. In addition, most of social networks are microblogging websites, e.g. messages on Twitter, or \textit{tweets}, are restricted to 140 characters. As a consequence, we often have to deal with short texts. For all the above reasons, topic modeling on those kinds of datasets tends to be a daunting task. \\

Several works used standard topic models to model topics on short texts, but observed a severe degradation in the performances of those algorithms \cite{mehrotra2013improving, yan2013biterm, likhitha2019detailed}.  This is mainly due to the lack of co-occurrence information in short documents, indeed, short texts typically only include a few words. In order to overcome this problem of sparsity, new models such as Dirichlet Multinomial Mixture (DMM) \cite{yin2014dirichlet}, have been introduced \cite{cheng2014btm, qiang2020short}. DMM is a simplication of the LDA model, in which each document is supposed to have been generated from only one topic instead of a mixture of topics. Recent studies suggest that DMM is more adapted to short texts than LDA and PLSA.  Several research papers also propose models that incorporate \textit{metadata} such as information about the author \cite{rosen2012author} or token embeddings \cite{gao2019incorporating, li2018lda},  but are out of the scope of this paper. \\

Prior to this work, we collected textual data from two data sources: \textit{tweets} from prominent personalities  of the Bitcoin community and messages posted on the specialized forum BitcoinTalk.org. The contribution of this article is to train three different models: LDA \cite{blei2003latent},  prodLDA \cite{srivastava2017autoencoding} and DMM \cite{yin2014dirichlet} on the dataset of \textit{tweets}, and to evaluate their performances on the Bitcoin Talk dataset. A model is said to be performing if the output topics are semantically coherent for humans. Since objective functions that are optimized by those models do not always correlate with human judgments of topic quality, we use several evaluation measures that have been explicitly developed to evaluate the coherence of models. As previously said, a trained topic model can be used to to find Euclidean representations of textual data. We finally show in this work that the representations found can be significant for predicting bitcoin's future returns.  

\section{Data}

\begin{figure*}[!ht]
    \centering
    \includegraphics[width=0.99\linewidth]{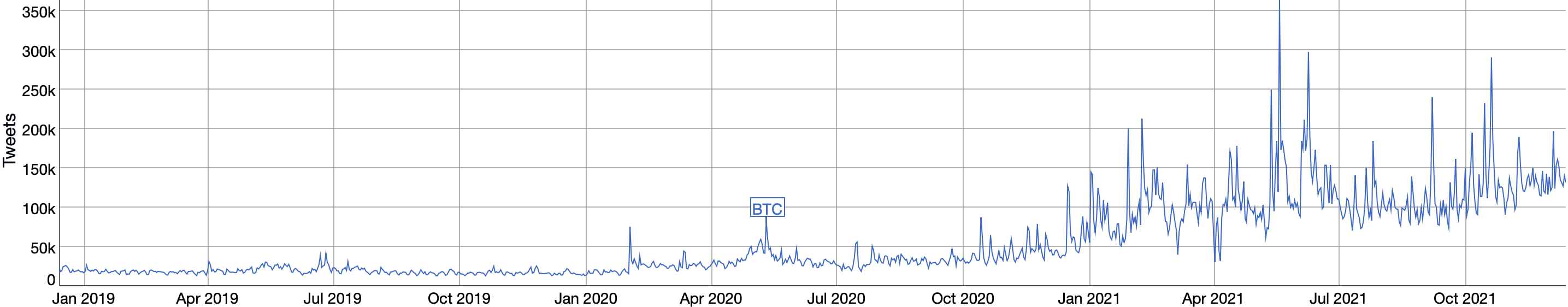}
    \caption{Number of tweets mentioning "btc" or "bitcoin" (source: \url{https://bitinfocharts.com}).}
    \label{fig:number_tweets}
\end{figure*}

\begin{figure*}[!htb]
\RawFloats
\begin{minipage}{.49\linewidth}
    \centering
    \includegraphics[width=0.95\columnwidth]{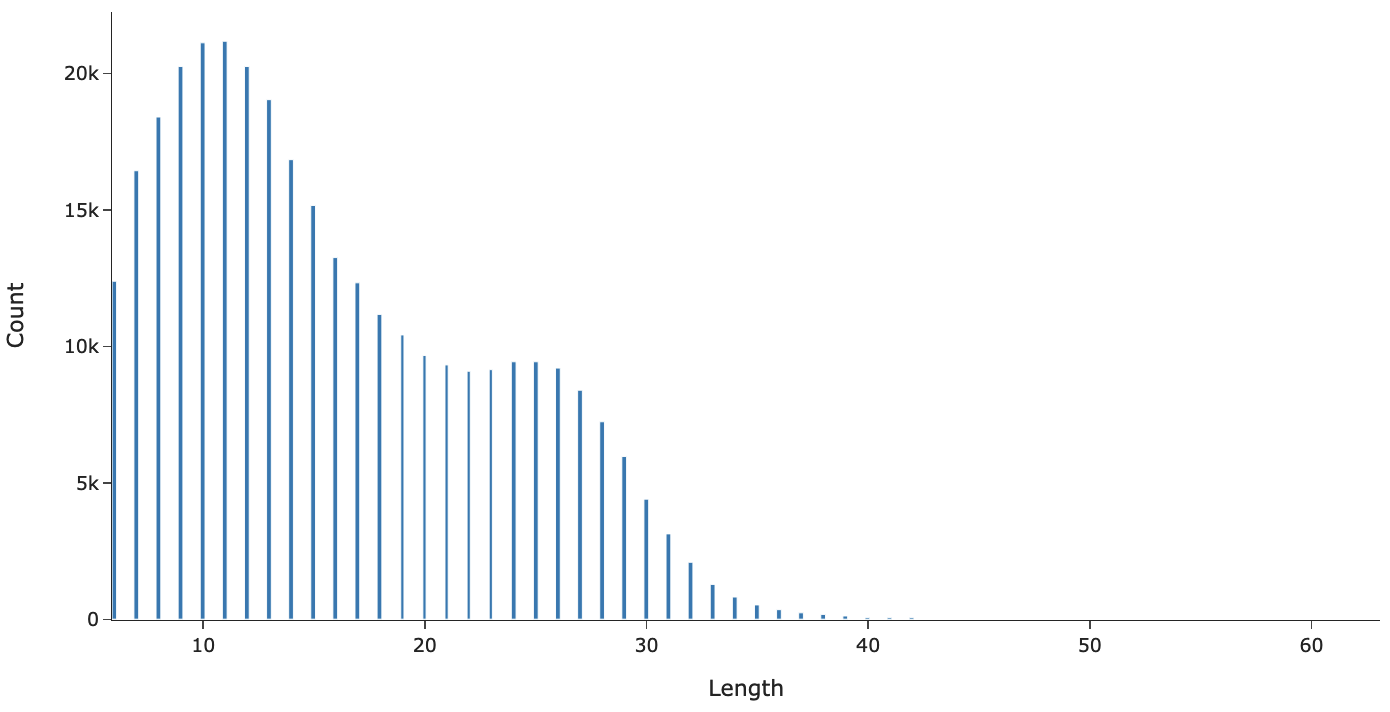}
    \caption{Twitter: histogram of length.}
    \label{fig:histo_twitter}
    \end{minipage}\hfill
    \begin{minipage}{.49\linewidth}
    \centering
    \includegraphics[width=0.95\columnwidth]{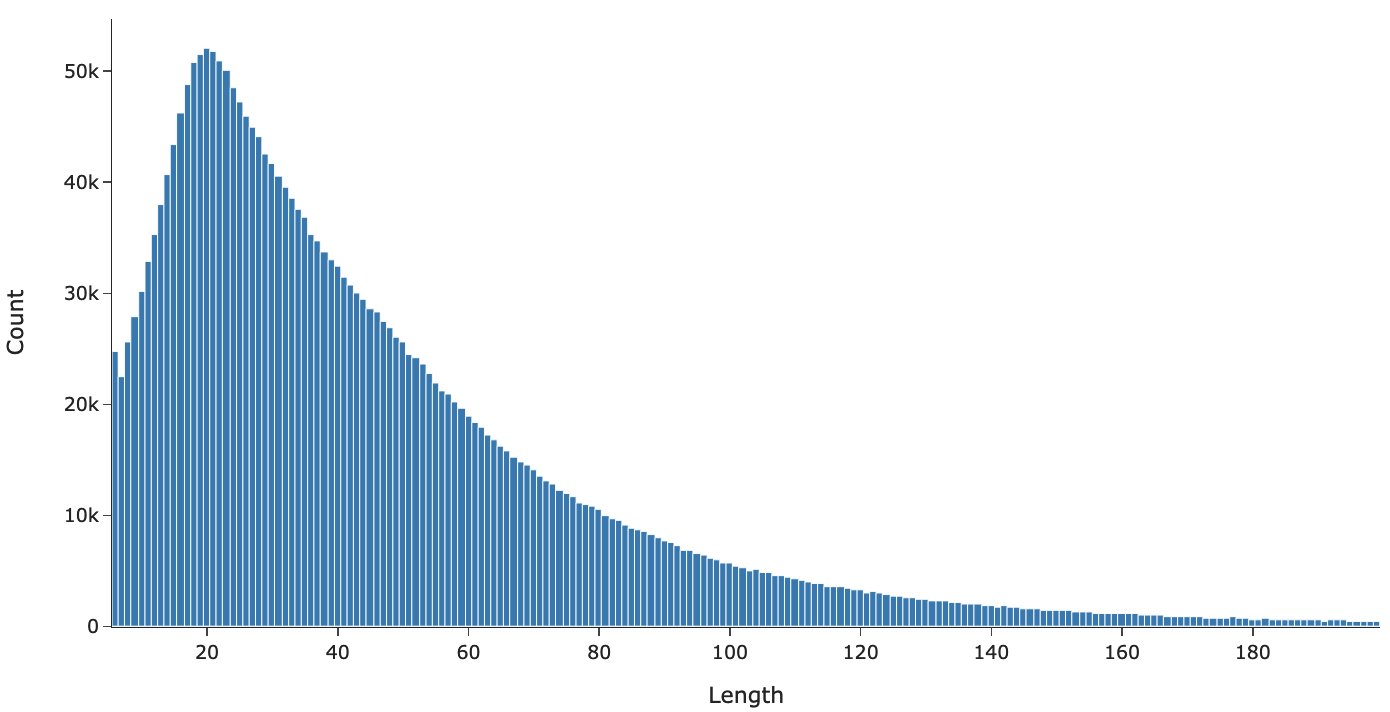}
    \caption{Bitcoin Talk: histogram of length.}
    \label{fig:histo_BT}
    \end{minipage}
\end{figure*}

Prior to this work, we collected two datasets of texts from social networks : first, a set of \textit{tweets} of prominent Twitter accounts in the cryptocurrency community and secondly, a set of messages posted on the forum Bitcoin Talk. \\

\paragraph{Twitter} Twitter (\url{https://twitter.com}) is a popular microblogging and social networking service on which people can post and interact with short messages, namely \textit{tweets}. Bitcoin community and more generally cryptocurrency communities are very established on Twitter. Figure \ref{fig:number_tweets} shows that the daily volume of tweets mentioning "bitcoin" has been increasing steadily since 2019. The number of tweets published about Bitcoin can culminate up to several hundreds of thousand each day. The huge quantity of tweets published each day makes it difficult to collect all of them. In addition, most of these tweets can be considered as spam \cite{perry2018effect}, or are not likely to have an influence on the propagation of news. We decide first to establish a list of prominent accounts of the Bitcoin community, and, then, to collect all tweets from these accounts. To this end, we collect all tweets mentioning "btc" or "bitcoin" and having generated more than 5000 \textit{retweets} from the 30th June 2017 to the 30th June 2021. Next, we filter out accounts that are present less than 5 times in order not to take into account accounts that were sporadically viral. Finally, we collect all tweets mentioning "btc" or "bitcoin" from the remaining accounts, this time without restriction on the number of retweets. \\

\paragraph{Bitcoin Talk} Bitcoin Talk (\url{https://bitcointalk.org}) is one the most popular online forum dedicated to cryptocurrency and blockchain technology. In particular, people can ask questions and discuss topics related to Bitcoin. The forum was even created by the creator of Bitcoin itself, Satoshi Nakamoto. There are several sections in which several thousands of discussions, or \textit{threads}, are available. We collect all messages published in the following sections : "Bitcoin Discussion", "Economics", "Gambling", "Speculation", "Securities", "Press", "Trading discussion" and "Legal", from the 22th November 2009 to the 30th Mai 2021. \\

We report in table \ref{tab:dataset_statistics} several statistics on both datasets.  In figures \ref{fig:histo_twitter} and \ref{fig:histo_BT} we plot histograms of the length of the collected tweets and messages on Bitcoin Talk, respectively. We observe that tweets are globally shorter than messages on Bitcoin Talk. The dataset \textbf{Twitter} will be used to train our models and the dataset \textbf{Bitcoin Talk} will be used to select the best model, as evaluation metrics are more reliable with longer texts. \\

\begin{table}
    \centering
    \begin{tabular}{c|ccc}
    \textbf{Dataset} & Documents & Authors & Threads  \\
    \hline
    \textbf{Twitter} & 328055 & 310 &  / \\
    \textbf{Bitcoin Talk} & 2552937 & 92193 & 46614 \\
    \end{tabular}
    \caption{Dataset statistics}
    \label{tab:dataset_statistics}
\end{table}

\section{Methods}

Notations:
\begin{itemize}
    \item $\mathcal{D}$ the training corpus
    \item $N_d \triangleq | \mathcal{D} |$ the number of documents
    \item $K \in \mathbb{N}$ the number of topics
    \item $V \in \mathbb{N}$ the vocabulary size
    \item $\Delta(n) \triangleq \{ \mathbf{p} \in \mathbb{R}^n, \ \mathbf{p} \geq 0, \ \sum_{i=1}^n p_i = 1\}$
    \item $\sigma$ the softmax function
    \item $\mathrm{diag}(V)$ is the diagonal matrix whose diagonal is equal to $V$
\end{itemize}

In this section, we will discuss some of the topic modeling methods that deals with words, documents and topics. All the studied models make the Naive Bayes assumption, i.e. words in a document are generated independently.

\label{sec:models}
\subsection{Dirichlet Multinomial Mixture}

\subsubsection{Model}

\begin{figure}
    \centering
\includegraphics[width=0.9\columnwidth]{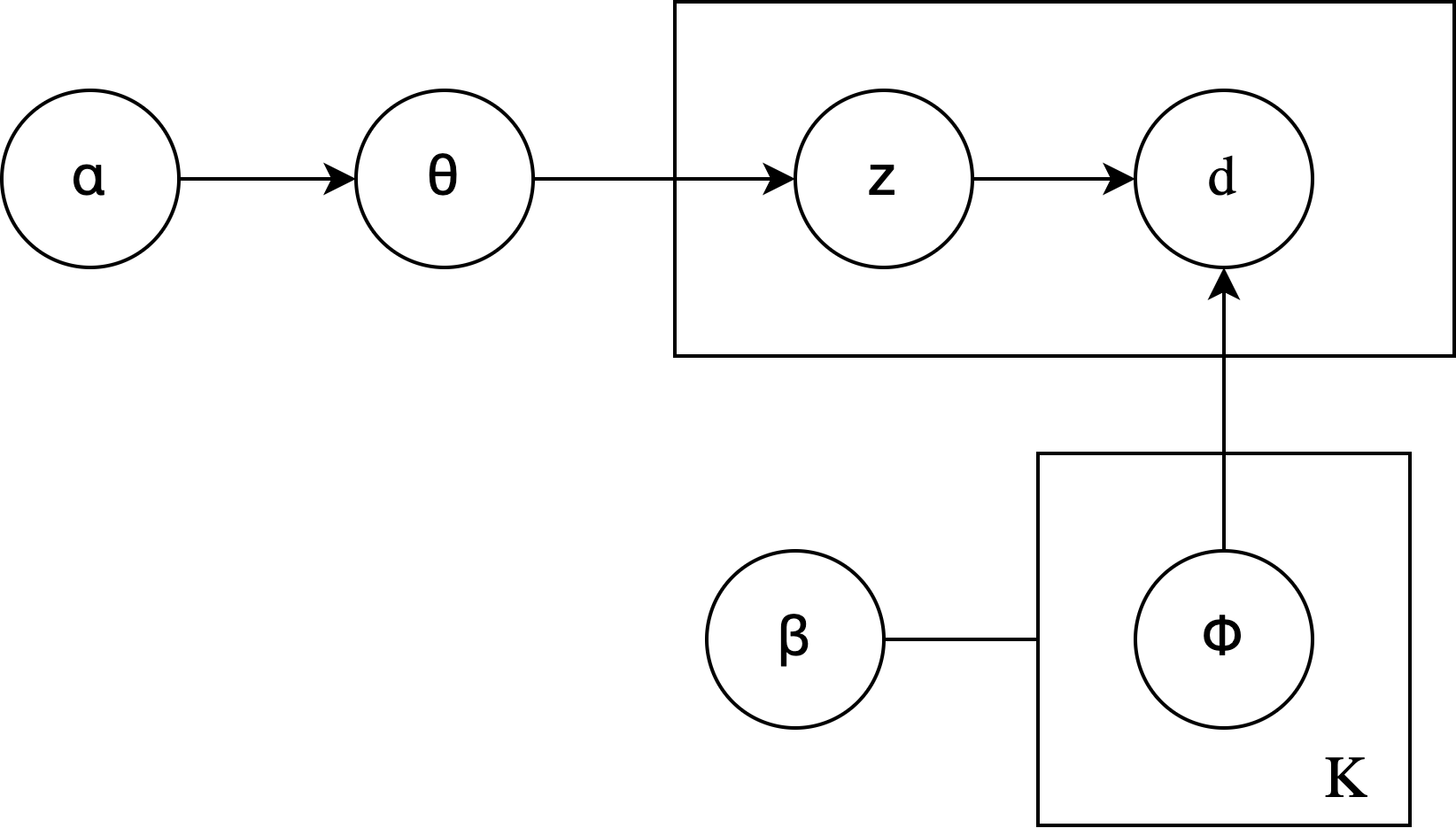}
\caption{Graphical Model of DMM.}
\label{fig:graphical_model_DMM}
\end{figure}

\emph{Dirichlet Multinomial Mixture} (DMM) is a generative probabilistic model in which each document covers an unique topic.  This model was introduced by \citeauthor{yin2014dirichlet} in \cite{yin2014dirichlet}. It is parametrized by two vectors $\boldsymbol{\alpha} \in (\mathbb{R}^{+*})^K$  and $\boldsymbol{\beta} \in (\mathbb{R}^{+*})^V$ defining two Dirichlet priors. We suppose that these priors are symmetric, i.e. $\boldsymbol{\alpha} \propto \overrightarrow{1}$ and $\boldsymbol{\beta} \propto \overrightarrow{1}$, this assumption implies that all topics and words are equally important in the beginning. In the following, we will denote by $\alpha$ and $\beta$ the real numbers satisfying $\boldsymbol{\alpha} = \alpha \cdot \overrightarrow{1}$ and $\boldsymbol{\beta} = \beta \cdot \overrightarrow{1}$.  Hyper-parameters $\alpha$ and $\beta$ can be here easily interpreted through an analogy, see the reference \cite{yin2014dirichlet} for further details.  A corpus is characterized by a topic probability distribution $\boldsymbol{\theta} \in \Delta(K) \sim \mathrm{Dirichlet}(\boldsymbol{\alpha})$, $\theta_t$ is the asymptotic proportion of documents covering topic $t$. In addition, each topic $t$ is characterized by a probability vector $\boldsymbol{\phi_t} \in \Delta(V)$ over the vocabulary, the $\boldsymbol{\phi_1}, ..., \boldsymbol{\phi_K}$ are i.i.d samples from $\mathrm{Dirichlet}(\boldsymbol{\beta})$. The generative process of a new document $d$ is defined as follows: first a topic $z$ is sampled from $\mathrm{Multinomial}(1, \boldsymbol{\theta})$, then $N$ words $\{w_1, ..., w_N\}$\footnote{Note that this model relies on the bag-of-words hyptothesis : the ordering of the words is not taken into account.} are i.i.d sampled from $\mathrm{Multinomial}(1, \boldsymbol{\phi_z})$. This graphical model is represented on figure \ref{fig:graphical_model_DMM} and the generative process on algorithm \ref{algo:dmm}.

\begin{algorithm}
\KwData{$N \in \mathbb{N}$}
\KwResult{Document $d$}
$z \sim \mathrm{Multinomial}(1; \boldsymbol{\alpha})$\;
$d \sim \mathrm{Multinomial}(N; \boldsymbol{\phi_z})$\;
\vspace*{0.3cm}
\caption{DMM: generative process of new documents.}
\label{algo:dmm}
\end{algorithm}

The joint probability of this graphical model therefore satisfies :

\begin{equation}
\small
        p(d, \alpha, \beta, z, \boldsymbol{\phi}, \boldsymbol{\theta}) = p(\alpha) p(\boldsymbol{\theta}|\alpha) p(z | \boldsymbol{\theta}) p(d|z,\boldsymbol{\phi}) p(\boldsymbol{\phi} | \beta) p(\beta) 
\end{equation}

where the probability of $d$ conditioned on $\boldsymbol{\theta}$ and $\boldsymbol{\phi}$ is:

\begin{equation}
\small
    \begin{split}
        p(d | \boldsymbol{\theta}, \boldsymbol{\phi}) &= \sum_z p(d | z, \boldsymbol{\theta}, \boldsymbol{\phi}) p(z |  \boldsymbol{\theta}, \boldsymbol{\phi}) \\
        &= \sum_z p(d | z, \boldsymbol{\phi}) p(z |  \boldsymbol{\theta}) \\
        &= \sum_z \left( \prod_{w \in d } p(w | z, \boldsymbol{\phi}) \right) \theta_z \\
        &= \sum_z  \left( \prod_{w \in d } \phi_{z, w} \right) \theta_z \\
    \end{split}
    \label{eq:dmm_likelihood}
\end{equation}

\vspace{0.5cm}
\subsubsection{Inference}
\label{sec:dmm_inference}
A corpus can be explored by examining the posterior distribution of $\boldsymbol{\phi}$, $\boldsymbol{\theta}$ and assignments $\mathbf{z}$ conditioned on the train documents. However the posterior cannot be computed directly \cite{yin2014dirichlet}. We use the collapsed Gibbs sampling algorithm, a MCMC algorithm, introduced by \citeauthor{yin2014dirichlet} to approximate the posterior given the train corpus. $\alpha$ and $\beta$ are fixed at the beginning during the initialization.  This method first initializes all topic labels randomly, next it iterates several times over the documents. For each document $d$, its new label is updated according to the conditional distribution $p(z_d | d, \{z_{d^{'}}\}_{d^{'} \in \mathcal{D} - \{d\} }, \alpha, \beta)$.

\vspace{0.5cm}
\subsubsection{Estimation of the topic-word matrix}
\label{seq:estimation_dmm}

At the end of the training phase, each document $d$ has been assigned a label $z_d$. We introduce the following notations:
\begin{itemize}
    \item $\mathcal{Z} \triangleq \{z_d\}_{d \in \mathcal{D}}$  the set of topic assignments, 
    \item $\mathcal{D}_i \triangleq \{d \in \mathcal{D}, \ z_d = i\}$ the set of documents covering topic $i$,
    \item $T_i \triangleq \sum_{d \in \mathcal{D}_i} |d|$ the total number of words of all documents covering topic $i$,
    \item $n^{T}_i \triangleq | \mathcal{D}_i|$ the number of documents covering topic $i$
    \item $n_{i,w}^{W}$ the number of occurence of $w$ in $\mathcal{D}_i$,
\end{itemize}

The posterior probability function of $\boldsymbol{\phi_i}$ given the documents and the inferred labels can be derived as follows:

\begin{equation} \label{eq:dmm_posterior_phi}
\small
\begin{split}
p(\boldsymbol{\phi_i} | \mathcal{Z}, \mathcal{D}, \alpha, \beta)
 & \overset{(B)}{\propto} p(\mathcal{D} | \boldsymbol{\phi_i}, \mathcal{Z}, \alpha, \beta) p(\boldsymbol{\phi_i} | \mathcal{Z}, \alpha, \beta) \\
 & \overset{(AG)}{\propto}  p(\mathcal{D}_i| \boldsymbol{\phi_i}, \mathcal{Z}) p(\boldsymbol{\boldsymbol{\phi_i}} |\beta) \\
 & \overset{(AD)}{\propto} \mathrm{Multinomial}(\mathcal{D}_i; T_i, \boldsymbol{\phi_i}) \mathrm{Dirichlet}(\boldsymbol{\phi_i}; \boldsymbol{\beta}) \\
 & \overset{(A1)}{\propto} \mathrm{Dirichlet}(\boldsymbol{\phi_i}; V, \boldsymbol{\beta} + \boldsymbol{n^{W}_{i}})
\end{split}
\end{equation}

where we use the following arguments : $(B)$ Bayes' theorem, $(AG)$ structure of the graphical model, $(AD)$ hypothesis of DMM about conditional distributions and $(A1)$ the fact that Dirichlet distribution is conjugate to the multinomial distribution. Finally, $\boldsymbol{\phi_i}$ can be estimated by the mean $\boldsymbol{\hat{\phi}_i}$  of the posterior:

\begin{equation}
\small
    \boldsymbol{\hat{\phi}_i}= \left( \frac{\beta + n^{W}_{i, w}}{T_i + V \beta} \right)_{w=1, ..., V}
    \label{eq:estimation_phi}
\end{equation}

By using the same kind of arguments, 

\begin{equation} \label{eq:dmm_posterior_theta}
\small
\begin{split}
p(\boldsymbol{\theta} | \mathcal{Z}, \mathcal{D}, \alpha, \beta)
 & \propto p(\mathcal{Z}| \boldsymbol{\theta}) p(\boldsymbol{\boldsymbol{\theta}} |\alpha) \\
 & \propto \mathrm{Multinomial}(\mathcal{Z}; N_d, \boldsymbol{\theta}) \mathrm{Dirichlet}(\boldsymbol{\theta}; \boldsymbol{\alpha}) \\
 & \propto \mathrm{Dirichlet}(\boldsymbol{\theta}; K, \boldsymbol{\alpha} + \boldsymbol{n^{T}})
\end{split}
\end{equation}

$\boldsymbol{\theta}$ is approximated by $\boldsymbol{\hat{\theta}}$ the posterior mean: 

\begin{equation}
\small
    \boldsymbol{\hat{\theta}} = \left( \frac{\alpha + n^{T}_{i}}{N + K \alpha} \right)_{i=1, ..., K}
    \label{eq:estimation_theta}
\end{equation}

\subsection{Latent Dirichlet Allocation}

\subsubsection{Model}

\begin{figure}
    \centering
    \includegraphics[width=0.9\columnwidth]{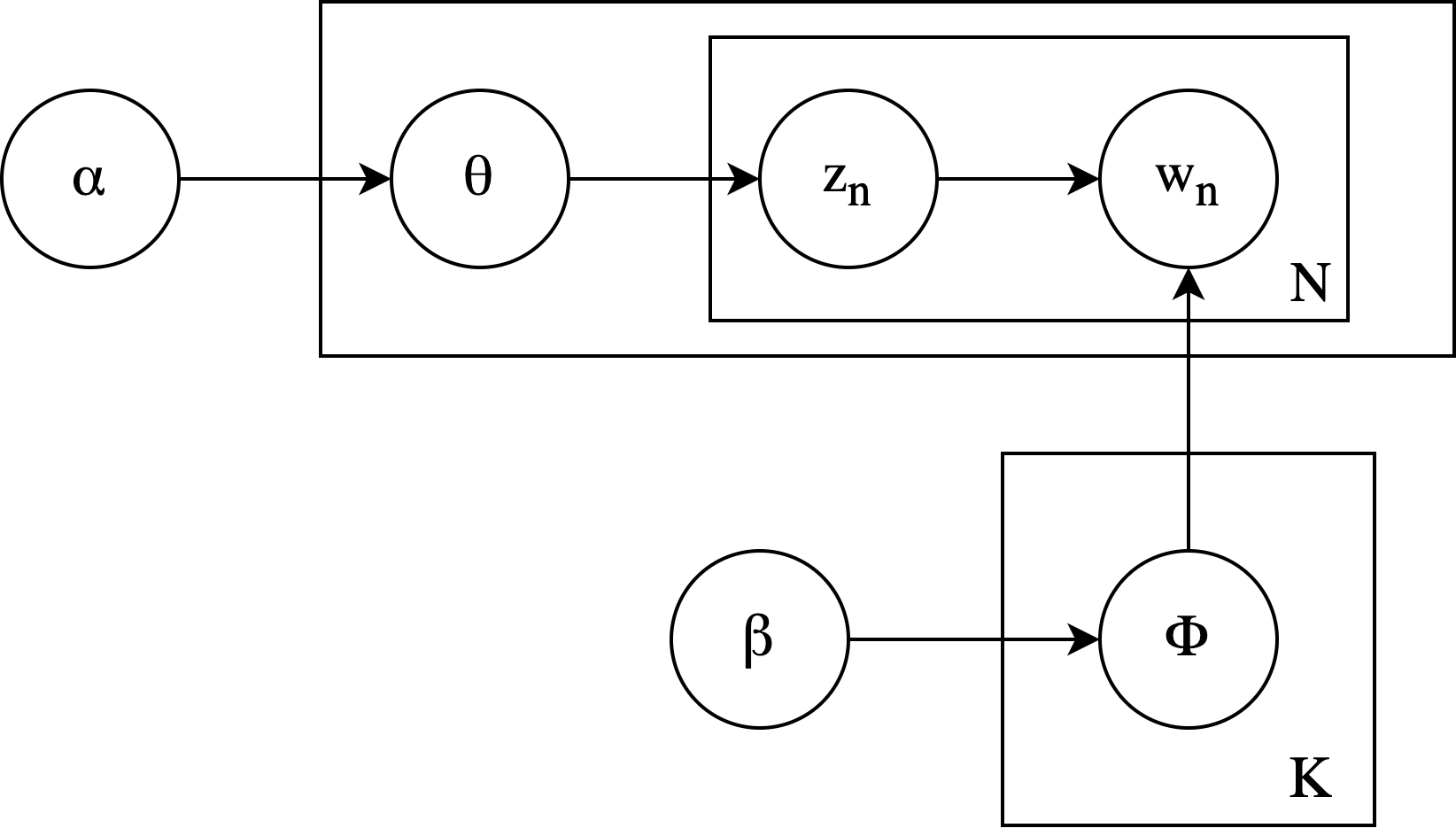}
    \caption{Graphical Model of LDA}
    \label{fig:graphical_model_LDA}
\end{figure}

\emph{Latent Dirichlet Allocation} (LDA) is a generative probabilistic model introduced by \citeauthor{blei2003latent}\cite{blei2003latent} in which each document is represented as a mixture of topics in contrary to DMM for which only one topic is covered.  As the DMM model, LDA is parametrized by $\boldsymbol{\alpha} \in (\mathbb{R}^{+*})^K$ and $\boldsymbol{\beta} \in (\mathbb{R}^{+*})^V$ defining two Dirichlet priors. Topics are characterized by $\{\boldsymbol{\phi_1}, ..., \boldsymbol{\phi_K}\}$ i.i.d sampled from $\mathrm{Dirichlet}(\boldsymbol{\beta})$. For generating a new document $d$, a topic distribution $\boldsymbol{\theta}$ is sampled from $\mathrm{Dirichlet}(\boldsymbol{\alpha})$. Then, for each word position $n$, we sample a topic $z_n$ from $\mathrm{Multinomial}(1, \boldsymbol{\theta})$ and a word $w_n$ according to $\mathrm{Multinomial}(1, \boldsymbol{\phi_{z_n}}$). As in DMM, we suppose that the Dirichlet priors are symmetric. This graphical model is represented on figure \ref{fig:graphical_model_LDA} and the generative process on algorithm \ref{algo:lda}.

\begin{algorithm}
\KwData{$N \in \mathbb{N}$}
\KwResult{Document $d = \{w_1, ..., w_N\}$}
$\boldsymbol{\theta} \sim \mathrm{Dirichlet}(\boldsymbol{\alpha})$\;
\For{$n=1,..., N$}{
$z_n \sim \mathrm{Multinomial}(1; \boldsymbol{\theta})$\;
$w_n \sim \mathrm{Multinomial}(1; \boldsymbol{\phi_{z_n}})$\;
}
\vspace*{0.3cm}
\caption{LDA : generative process of a new document.}
\label{algo:lda}
\end{algorithm}

The joint probability of this graphical model therefore satisfies :
  
\begin{equation}
\small
p(d, \alpha, \beta, \mathbf{z}, \boldsymbol{\phi}, \boldsymbol{\theta}) = p(\alpha) p(\boldsymbol{\theta} | \alpha) p(d | \mathbf{z}, \boldsymbol{\phi})  p(\mathbf{z} | \boldsymbol{\theta})  p(\boldsymbol{\phi} | \beta) p (\beta) 
\end{equation}

Furthermore, 

$$ p(\mathbf{z} | \boldsymbol{\theta}) p(d | \mathbf{z}, \boldsymbol{\phi}) = \prod_{i=1}^{N} p(w_n | z_n, \boldsymbol{\phi}) p(z_n | \boldsymbol{\theta})$$

The probability of $d$ conditioned on ${\alpha}$ and $\boldsymbol{\phi}$ is intractable, indeed:

\begin{equation}
\small
    \begin{split}
        p(d | \alpha, \boldsymbol{\phi}) &= \int_{\Delta(K)} p(d | \alpha, \boldsymbol{\theta}, \boldsymbol{\phi}) p(\boldsymbol{\theta} | \alpha, \boldsymbol{\phi}) \mathrm{d} \boldsymbol{\theta} \\
        &= \int_{\Delta(K)} p(d | \boldsymbol{\theta}, \boldsymbol{\phi}) p(\boldsymbol{\theta} | \alpha) \mathrm{d} \boldsymbol{\theta} \\
        &= \int_{\Delta(K)} \left( \prod_{w \in d} \sum_z p(w | z, \boldsymbol{\phi_z}) p(z| \boldsymbol{\theta}) \right) p(\boldsymbol{\theta} | \alpha) \mathrm{d} \boldsymbol{\theta}
    \end{split}
\end{equation}

\vspace{0.5cm}
\subsubsection{Inference}

As in DMM, the posterior distribution is intractable. The variational Bayes algorithm is traditionally used to this end. The true posterior is approximated by a simpler distribution $q(\boldsymbol{\phi}, \boldsymbol{\theta}, \boldsymbol{z} | \boldsymbol{\chi})$ with $\boldsymbol{\chi}$ a set of free parameters. Parameters $\boldsymbol{\chi}$ are optimized to maximise the Evidence Lower Bound, or shortly ELBO, quantity (equation \ref{eq:elbo_lda}). 
\label{sec:lda_inference}

\begin{equation}
\small
\begin{split}
     \mathrm{ELBO}(\mathcal{D}, \boldsymbol{\chi}) & = \mathbb{E}_{q}(\log(p(\mathcal{D}, \boldsymbol{z}, \boldsymbol{\theta}, \boldsymbol{\phi} | \alpha, \beta))) \\
     &   - \mathbb{E}_q(\log(q(\boldsymbol{z}, \boldsymbol{\theta}, \boldsymbol{\phi} | \boldsymbol{\chi}))) 
\end{split}
\label{eq:elbo_lda}
\end{equation}

As its name suggests, the ELBO is a lower bound of the loglikelihood of the observed data, i.e. the evidence: 

\begin{equation}
\small
    \log (p ( \mathcal{D} | \alpha, \beta) )
\end{equation}

In particular we use the online method introduced by \citeauthor{hoffman2010online} \cite{hoffman2010online}.

\vspace{0.5cm}
\subsubsection{Estimation of the topic-word matrix}

The online variational Bayes algorithm of \citeauthor{hoffman2010online} allows to approximate the posterior of $\boldsymbol{\phi}$ conditioned on the corpus. $\boldsymbol{\phi}$ is finally estimated by the posterior mean.

\subsection{ProdLDA}

\subsubsection{Model}

First, it is worth noticing that topic assignments $\mathbf{z}$ in LDA can be collapsed by mixing the multinomial distributions:

\begin{equation}
\small
    \sum_z p(w | z, \boldsymbol{\phi}) p(z, \boldsymbol{\theta}) = \sum_z \phi_{z, w} \theta_w = (\boldsymbol{\phi} \times \boldsymbol{\theta})_{w}
\end{equation}

Thus, 

\begin{equation}
\small
    p(w | \boldsymbol{\phi}, \boldsymbol{\theta}) = \mathrm{Multinomial}(w; 1, \boldsymbol{\phi} \times \boldsymbol{\theta})
\end{equation}

The ProdLDA model is very similar to the LDA model. In ProdLDA, rows of the topic-word distribution matrix are not sampled from a Dirichlet distribution. Moreover, rows of $\boldsymbol{\phi}$ are not even constrained to live in $\Delta(V)$. Multinomial distributions are first mixed before being normalized into a distribution, that is:

\begin{equation}
\small
\begin{split}
    p(w | \boldsymbol{\phi}, \boldsymbol{\theta}) &= \mathrm{Multinomial}(w; 1, \sigma(\boldsymbol{\phi} \times \boldsymbol{\theta})) \\
    & \propto \prod_{z} p(w | z, \boldsymbol{\phi_z})^{\theta_z}
\end{split}
\label{eq:model_prodLDA}
\end{equation}

As a result, $w | \boldsymbol{\theta}, \boldsymbol{\phi}$ is a weighted product of expert instead of a mixture of multinomial. 

\vspace{0.5cm}
\subsubsection{Inference}

Computing the posterior is here again intractable. As for LDA,  we use a variational Bayes method in order to approximate the posterior distribution $p(\boldsymbol{\theta}, \boldsymbol{\phi} | \mathcal{D}, \alpha)$. \citeauthor{srivastava2017autoencoding} \cite{srivastava2017autoencoding} propose to use a \textit{Variational AutoEncoder} (VAE), a \textit{deep neural network}, to this end. A Variational AutoEncoder is composed of two successive modules : first the \textit{encoder}, or \textit{inference network} , and then the \textit{decoder}. A document $d$ is fed to the encoder which outputs an approximation $q$ of the posterior distribution $p(\boldsymbol{\theta} | d)$. $q(\boldsymbol{\theta} | d)$ belongs then to a parametric family of distributions, and is parametrized by $\boldsymbol{\chi}$, the free parameters of the encoder. Next, a sample $\boldsymbol{\theta_s}$ of $q(\boldsymbol{\theta} | d, \chi)$ is fed to the decoder, which outputs a probability vector $p(w | \boldsymbol{\theta_s}, \boldsymbol{\phi}) $ over the vocabulary (equation \ref{eq:decoder}). 

\begin{equation}
\small
   p(w | \boldsymbol{\theta_s}, \boldsymbol{\phi}) = \sigma(\boldsymbol{\phi} \times \boldsymbol{\theta_s})
    \label{eq:decoder}
\end{equation}

The decoder is thus composed of a linear layer without bias and a Softmax layer. \citeauthor{srivastava2017autoencoding} derive from the ELBO of equation \ref{eq:elbo_lda} an objective function $L$ (equation \ref{eq:ELBO_prodLDA}) to minimize. The parameter of the model $\{\boldsymbol{\chi}, \boldsymbol{\phi}\}$ are then optimized by a standard gradient descent algorithm. 

\begin{equation}
\small
\begin{split}
 L(\boldsymbol{\chi}, \boldsymbol{\phi} | d, \alpha) &=  \underbrace{D_{KL}( q(\boldsymbol{\theta} | \boldsymbol{\chi}, d) || p(\boldsymbol{\theta} | \alpha) )}_\text{(1)} \\  &+ \underbrace{\mathbb{E}_{ \boldsymbol{\theta_s} \sim q(\boldsymbol{\theta} | \boldsymbol{\chi}, d)} (- \log ( p(d | \boldsymbol{\theta_s}, \boldsymbol{\phi}))}_\text{(2)}
\end{split}
\label{eq:ELBO_prodLDA}
\end{equation}

\begin{figure}
    \centering
    \includegraphics[width=0.9\columnwidth]{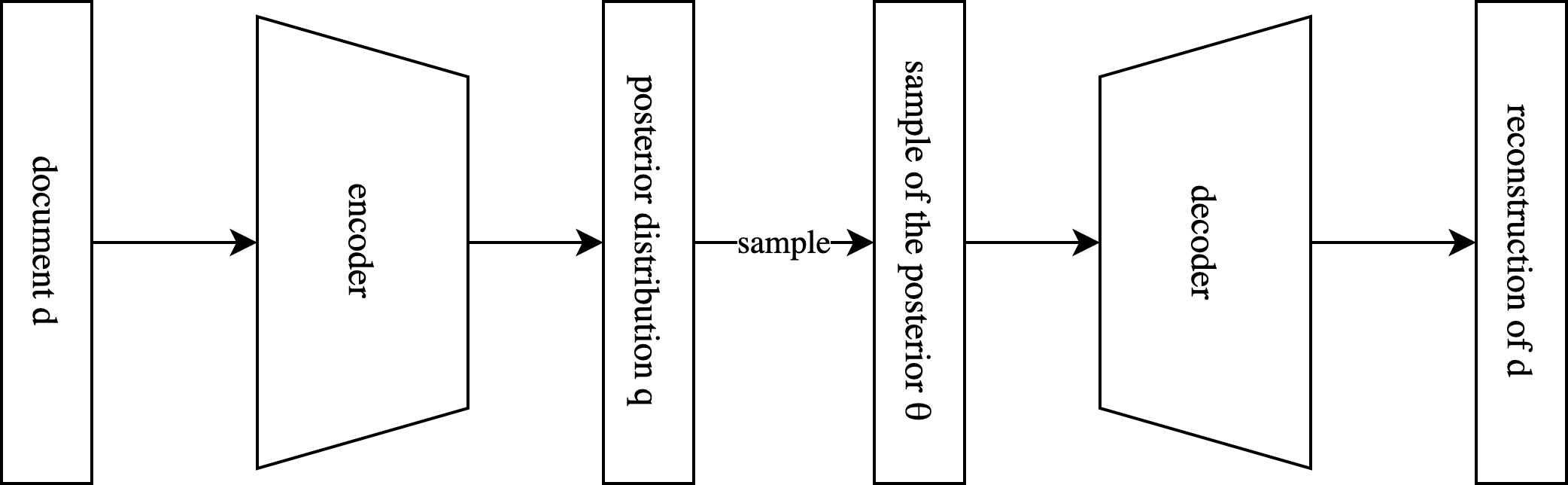}
    \caption{ProdLDA: Variational AutoEncoder}
    \label{fig:prodLDA_VAE}
\end{figure}

 The term (1) of equation \ref{eq:ELBO_prodLDA} is a regularization term which measures the difference between the approximate posterior distribution $q(\boldsymbol{\theta} | \boldsymbol{\chi}, d)$ and the prior distribution $p(\boldsymbol{\theta} | \alpha)$. As a consequence, the approximate posterior is forced to stay close to the prior.  The term (2) of equation \ref{eq:ELBO_prodLDA} is a reconstruction error which aims to maximize the probability of the input document in our model.

 The choice of the parametric family for $q$, i.e. the encoder, is crucial for two reasons:
 \begin{itemize}
     \item to derive a closed form of (1) in equation \ref{eq:ELBO_prodLDA}, 
     \item to use the traditional \textit{Reparametrization Trick} (RT) to sample from $q(\boldsymbol{\theta} | \boldsymbol{\chi}, d)$,
 \end{itemize}
 
  Choosing a Dirichlet prior here is problematic because it does not allow to use the RT. To deal with this, \citeauthor{srivastava2017autoencoding} use a Laplace approximation of the Dirichlet prior:  $p(\boldsymbol{\theta} | \alpha)$ is approximated by  $\hat{p}(\boldsymbol{\theta} | \boldsymbol{\tilde{\mu}}, \boldsymbol{\tilde{\Sigma}}) = \mathrm{LogNormal}(\boldsymbol{\theta}; \boldsymbol{\tilde{\mu}}, \boldsymbol{\tilde{\Sigma}})$, where $\boldsymbol{\tilde{\mu}} \in \mathbb{R}^K$ and $\boldsymbol{\tilde{\Sigma}} \in (\mathbb{R}^{+*})^K$ are computed from the parameter vector $\boldsymbol{\alpha}$ of the Dirichlet prior:

\begin{equation}
\small
    \tilde{\mu}_k = \log(\alpha_k) - \frac{1}{K} \sum_i \log \alpha_i
\end{equation}

\begin{equation}
\small
    \tilde{\Sigma}_{kk} = \frac{1}{\alpha_k} \left( 1 - \frac{2}{K} \right) + \frac{1}{K^2} \sum_i \frac{1}{\alpha_k}
\end{equation}

The posterior $q$ is also approximated by a logistic normal distribution with a diagonal covariance matrix, i.e. $q(\boldsymbol{\theta}| \boldsymbol{\chi}, d )  = \mathrm{LogNormal}(\boldsymbol{\theta} ; \boldsymbol{\mu}(d), \mathrm{diag}(\boldsymbol{\Sigma}(d)))$ where $\boldsymbol{\mu}(d) \in \mathbb{R}^K$ and $\boldsymbol{\Sigma}(d) \in (\mathbb{R}^{+*})^K$ are the outputs of the encoder. As a consequence, we can rewrite (1) of equation \ref{eq:ELBO_prodLDA} as:

\begin{equation}
\small
\begin{split}
D_{KL}(q(\boldsymbol{\theta} | \boldsymbol{\chi}, d) || \hat{p}(\boldsymbol{\theta} | \boldsymbol{\alpha}) ) = \frac{1}{2} \bigg ( \mathrm{tr}(\tilde{\boldsymbol{\Sigma}}^{-1} \boldsymbol{\Sigma}( d)) \\ + (\boldsymbol{\tilde{\mu}} - \boldsymbol{\mu}(d))^T \boldsymbol{\Sigma}^{-1} (\boldsymbol{\tilde{\mu}} - \boldsymbol{\mu}(d)) - K + \log \left(  \frac{|\tilde{\boldsymbol{\Sigma}}|}{|\boldsymbol{\Sigma}(d)|} \right)  \bigg ) 
\end{split}
\label{eq:prodlda_dkl}
\end{equation}

RT can be used with logistic normal distribution, indeed if  $\boldsymbol{\epsilon}$ is sampled from  $\mathrm{Normal}(0, \boldsymbol{I})$, then:

\begin{equation}
\small
     \sigma \left(\boldsymbol{\mu} + \sqrt{\mathrm{diag}(\boldsymbol{\Sigma})} \boldsymbol{\epsilon} \right)
     \sim \mathrm{LogitNormal}(\boldsymbol{\theta}; \boldsymbol{\mu}, \mathrm{diag}(\boldsymbol{\Sigma})) 
    \label{eq:prodlda_mc}
\end{equation}

Finally, the expected value of term (2) of equation \ref{eq:ELBO_prodLDA} is approximated by a Monte Carlo estimator with one sample:

\begin{equation}
\small
\begin{split}
    \mathbb{E}_{\boldsymbol{\theta} \sim q(\boldsymbol{\chi}, d)} \left( - p(d | \boldsymbol{\theta}, \boldsymbol{\phi}) \right) \approx  \\- \sum_{w} 1_{w \in d} \log \left( \sigma \left( \boldsymbol{\phi} \times \sigma \left(\boldsymbol{\mu}(d) + \sqrt{\boldsymbol{\Sigma}(d)} \boldsymbol{\epsilon} \right) \right) \right),  
\end{split}
\label{eq:reconstruction_decoder}
\end{equation}
where $\boldsymbol{\epsilon} \sim \mathrm{Normal}(0, \boldsymbol{I})$
\vspace{0.5cm}

\subsubsection{Estimation of the topic-word matrix}

$\boldsymbol{\phi_i}$ is obtained by applying the Softmax function to the $i$-th row of the weight matrix $\boldsymbol{\phi}$ of the decoder.

\section{Evaluation metrics}

In the different models, a topic $i$ is defined by a distribution vector $\boldsymbol{\phi_i}$. A topic can thus be described its the most frequent words, or equivalently words with the highest probability. Intuitively, a subject is good if its top words are semantically coherent, that is, these words are meaningful enough so that a human can easily understand the main theme. Semantic coherence is subject to human subjectivity and judgement, and is therefore difficult to evaluate automatically. In this section, we present some metrics that are traditionally used to evaluate and select topic models.  
 
\subsection{Human judgement}
\label{sec:human_judgement}

As previously said, even if it may differ from one person to another, human judgment is the gold standard. To this end, human experts usually explore the set of top words by frequency in each topic. However, it may not be the best reduction technique for visualizing the topics and  evaluating their quality. Indeed,  overrepresented words such as "btc" or "bitcoin" are very likely to be top words for all topics, but do not give much information about the semantic of the topic. Arising out of this concern, \citeauthor{sievert2014ldavis} \cite{sievert2014ldavis} introduce the \textit{relevance} score $r$. The relevance $r_{\lambda}(w, i)$ of word $w$ for topic $i$ is computed as follow: 

\begin{equation}
\small
    r_{\lambda}(w, i) = \lambda \log( \phi_{i,w}) + (1-\lambda) \log \left( \frac{\phi_{w, i}}{p_w}\right)
    \label{eq:relevance}
\end{equation}

where $\lambda \in [0, 1]$ and $p_w$ is the empirical probability of $w$ in the corpus. If $\lambda = 1$, ordering by topic-relevance is equivalent to ordering by topic-frequency. In contrary, if $\lambda = 0$, importance is given to words essentially only present in this topic. $\lambda$ is therefore a tradeoff parameter, \citeauthor{sievert2014ldavis} recommend to set $\lambda$ equal to 0.3.

\subsection{Perplexity}

Let $\mathcal{D}^{(\mathrm{test})}$ a set of $N^{(\mathrm{test})}$ new  documents $\{d_1, d_2, ..., d_{N^{(\mathrm{test})}}\}$ generated with the same model as $\mathcal{D}$ was. Let $\mathcal{L}$ be the log likelihood function of our generative model. We can measure the quality of our model by evaluating the likelihood of the new document set $\mathcal{D}^{(\mathrm{test})}$, or \textit{perplexity}. A high likelihood, or equivalently a high log-likelihood, would mean that our model is not surprised by the new documents, that is, the model has been quite good calibrated on the corpus $\mathcal{D}$. Since we supposed that documents are independently generated, we can decompose the log likelihood of the set into a sum of document log likelihood functions: 

\begin{equation}
\small
    \mathcal{L}(\mathcal{D}^{(\mathrm{test})} | \mathcal{D}) = \log p(\mathcal{D}^{(\mathrm{test})} | \mathcal{D}) = \sum_{i=1}^{N^{(\mathrm{test})}} \log p(d_i | \mathcal{D})
    \label{eq:perplexity}
\end{equation}
\label{sec:eval}

\emph{Perplexity} score is  defined as follows: 

\begin{equation}
\small
    \mathrm{perplexity}(\mathcal{D}^{(\mathrm{test})}) = \exp \left(- \frac{\mathcal{L}(\mathcal{D}^{(\mathrm{test})} | \mathcal{D})}{\sum_{i=1}^{N^{(\mathrm{test})}} |d_i|} \right)
\end{equation}

The lower the perplexity is, the better the model is assumed to be. It has been shown that perplexity is not a good metric for qualitative evaluation of topics \cite{chang2009reading}, it is not strongly correlated to human judgement, and sometimes even negatively correlated. However, we will use it to select hyper-parameters of our DMM and LDA models. \\

For DMM, the log likelihood of an unseen document can be computed with equation \ref{eq:dmm_likelihood} by remplacing $(\boldsymbol{\theta}, \boldsymbol{\phi})$ by $(\boldsymbol{\hat{\theta}}, \boldsymbol{\hat{\phi}})$ (equations \ref{eq:estimation_theta} and \ref{eq:estimation_phi}). 
$\mathcal{L}(\mathcal{D}^{(\mathrm{test})} | \mathcal{D})$ is intractable for LDA and ProdLDA. Following \citeauthor{blei2003latent} \cite{blei2003latent}, for variational methods we use the ELBO as a proxy for $\mathcal{L}(\mathcal{D}^{(\mathrm{test})} | \mathcal{D})$. For LDA, ELBO is computed by helding the topic matrix $\boldsymbol{\phi}$ fixed and fitting the free parameters over the topic distributions and topic assignments of new documents with the inference algorithm of  \cite{hoffman2010online}. For ProdLDA, new documents are fed to the VAE and ELBO is computed from equations \ref{eq:ELBO_prodLDA}, \ref{eq:prodlda_dkl} and \ref{eq:prodlda_mc}.

\subsection{Coherence measures}
\label{sec:coherence}

Each topic $i$ can be represented by its most frequent words according to $\boldsymbol{\phi_i}$. Given a topic $i$, we denote by $w_i^{(1)}, w_i^{(2)}, ..., w_i^{(V)}$ the words sorted by their frequencies in the descending order. Topic coherence measures take the set of $N$ top words of a topic and sum a confirmation measure over all word pairs. Once the confirmation measure is computed for all topics, they are aggregated into an unique coherence score. Several confirmation measures exist and we will study three of them in this section. UMass coherence is introduced by \citeauthor{mimno2011optimizing} \cite{mimno2011optimizing} and, given a set of document $\mathcal{D}$, UMass for topic $i$ is computed as follows:

\begin{equation}
\small
    C_{UMass, i} = \frac{2}{N(N-1)} \sum_{m=2}^{N} \sum_{n=1}^{m-1} \log \left( \frac{p_{\mathcal{D}}(w_i^{(m)}, w_i^{(n)}) + \epsilon}{p_{\mathcal{D}}(w^{(n)})} \right)
\end{equation}

where $\epsilon \approx 0^{+}$ is a small positive constant used to prevent numerical issues. The probability of a single word $p_{\mathcal{D}}(w)$ is defined as the number of documents of $\mathcal{D}$ in which the word $w$ occurs divided by the number of documents (equation \ref{eq:prob_single_word}). The joint probability of two words $p_{\mathcal{D}}(w_1, w_2)$ is estimated by the number of documents containing both words divided by the number of documents (equation \ref{eq:joint_prob_pair_word}).

\begin{equation}
\small
    p_{\mathcal{D}}(w) = \frac{\sum_{d \in \mathcal{D}} 1_{w \in d}}{|\mathcal{D}|}
    \label{eq:prob_single_word}
\end{equation}
\begin{equation}
\small
    p_{\mathcal{D}}(w_1, w_2) = \frac{\sum_{d \in \mathcal{D}} 1_{w_1 \in d} 1_{w_2 \in d} }{|\mathcal{D}|}
    \label{eq:joint_prob_pair_word}
\end{equation}

Let $\omega \in \mathbb{N}$ be a window size. We denote by $\mathcal{D}_{\omega}$ the set of documents obtained by sliding a window of size $\omega$ over all documents of $\mathcal{D}$. 
UCI coherence for topic $i$ uses a confirmation based on \emph{pointwise mutual information} (PMI) and is calculated by:

\begin{equation}
\small
    C_{UCI, i} = \frac{2}{N(N-1)} \sum_{m=1}^{N-1} \sum_{n=m+1}^N \mathrm{PMI} \left(w_i^{(m)}, w_i^{(n)} \right)
\end{equation}

with

\begin{equation}
\small
  \mathrm{PMI}\left(w_i^{(m)}, w_i^{(n)}\right)  = \log \left( \frac{p_{\mathcal{D}_{\omega}} \left( w_i^{(m)}, w_i^{(n)} \right) + \epsilon}{p_{\mathcal{D}_{\omega}} \left(w_i^{(m)}\right) p_{\mathcal{D}_{\omega}} \left(w_i^{(n)}\right)} \right)
  \label{eq:pmi}
\end{equation}

where probabilities in equation \ref{eq:pmi} are estimated on the set $\mathcal{D}_{\omega}$.  NPMI coherence is obtained from the formula of UCI coherence by remplacing the PMI with the \emph{normalized log-ratio measure} (NPMI) \cite{bouma2009normalized}. 

\begin{equation}
\small
    C_{NPMI, i} = \frac{2}{N(N-1)} \sum_{m=1}^{N-1} \sum_{n=m+1}^N \mathrm{NPMI} \left(w_i^{(m)}, w_i^{(n)} \right)
\end{equation}

\begin{equation}
\small
   NPMI(w_i^{(m)}, w_i^{(n)}) = \frac{PMI(w_i^{(m)}, w_j^{(n)})}{-\log(p_{\mathcal{D}_{\omega}}(w_i^{(m)}, w_i^{(n)}) + \epsilon)}
\end{equation}

Finally, coherence scores over all topics are aggregated by averaging. 

\begin{equation}
\small
    C_{\cdot} = \frac{1}{K} \sum_{i=1}^K C_{\cdot, i} 
\end{equation}

\citeauthor{roder2015exploring} \cite{roder2015exploring} show that these coherence measures are strongly positively correlated with human ratings.

\section{Materials and Results}

\subsection{Pre-processing}

The pre-processing of both datasets involves the following steps: 
\begin{enumerate}
    \item removal of non-Latin characters / emojis / urls, 
    \item removal of mentions of other users, typically "@user" in tweets,
    \item removal of English stop words and punctuation, 
    \item lemmatization
    \item tokenization
    \item[6.] (DMM) removal of duplicate tokens
\end{enumerate}

Finally, words that are present less than 100 times in the corpus are also removed. For steps 3 and 4,  we use the Python package \textsc{spaCy} (\url{https://spacy.io}) with the english pipeline (\url{https://spacy.io/models/en}). 

\subsection{Training}

In this section, we train the models presented in section \ref{sec:models} on the dataset \textbf{Twitter}. Since the optimal number of topics is not known \textit{a priori}, models are trained for a number of topics $K$ varying in $\{5, 10, 15, 20, 25, 30, 35, 40, 45, 50\}$. \\

The model DMM is trained with a Gibbs sampling algorithm presented in \ref{sec:dmm_inference}, in particular we use the implementation of the Python package \textsc{gsdmm} (\url{https://github.com/rwalk/gsdmm}). Hyper-parameters $\alpha$ and $\beta$ are optimized by a 5-fold cross-validated grid-search over the parameter grid $[0.01, 0.025, 0.05, 0.1, 0.2] \times [0.06, 0.1, 0.24]$. Perplexity (equation \ref{eq:perplexity}) is used to score the goodness-of-fit of the model. \\

The model LDA is trained with a variational online algorithm presented in section \ref{sec:lda_inference}. We use the popular Python package \textsc{Gensim} (\url{https://radimrehurek.com/gensim/}) for this purpose. Hyper-parameters $\kappa$ (decay) and $\tau_0$ (offset) are optimized by a 5-fold cross-validated grid search over the parameter grid $[0.6, 0.75, 0.9] \times [1, 64, 256]$. Perplexity, via its lower bound ELBO, is also used to select the best model.  \\

\begin{figure}
    \centering
    \includegraphics[width=0.7\columnwidth]{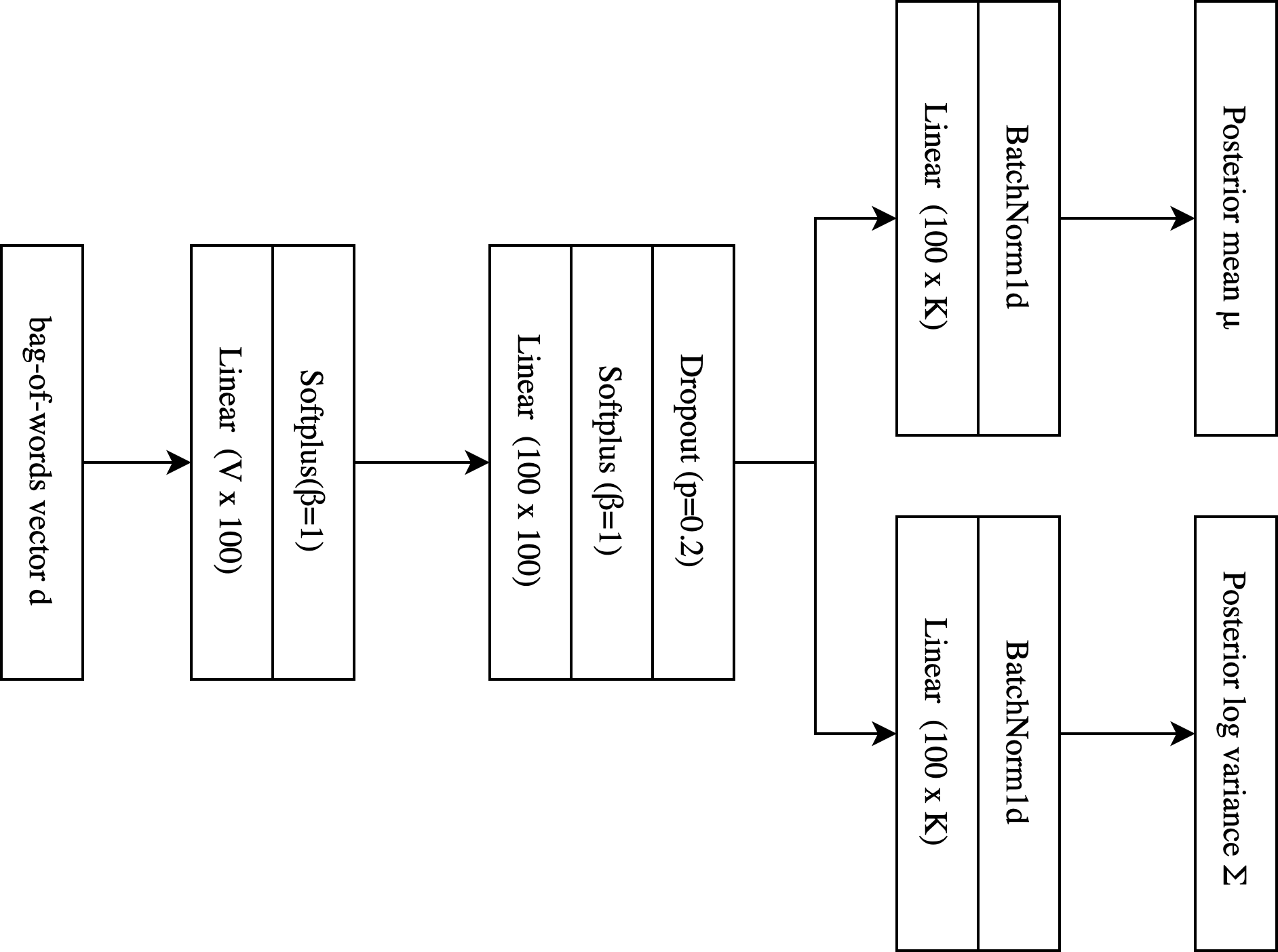}
    \caption{ProdLDA : inference network.}
    \label{fig:prodLDA_inference}
\end{figure}

\begin{figure}
    \centering
    \includegraphics[width=0.4\columnwidth]{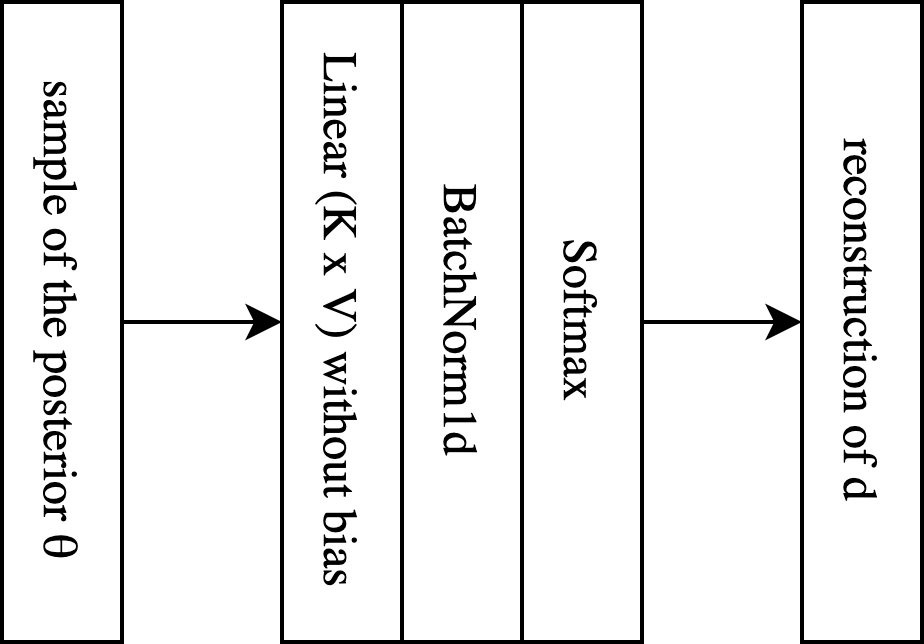}
    \caption{ProdLDA : decoder network.}
    \label{fig:prodLDA_decoder}
\end{figure}

The inference neural network of the VAE for ProdLDA is composed of three modules. The first module consists in two successive linear layers with Softplus activation, through which a document $d$ is summarized into a low-dimensional representation $\mathbf{z}$. Two parallel linear layers then transform $\mathbf{z}$ into the posterior mean $\boldsymbol{\mu}(d)$ and posterior log-variance $\log(\boldsymbol{\Sigma}(d))$. In accordance with equation \ref{eq:reconstruction_decoder}, the decoder is composed of a linear layer without bias and a $\mathrm{Softmax}$ layer. We use $\mathrm{BatchNorm}$ and $\mathrm{Dropout}$ layers to facilitate the training. The encoder and decoder are represented in figures \ref{fig:prodLDA_inference} and \ref{fig:prodLDA_decoder}, respectively. The training data is split into a train (70\%) and a validation (30\%) set. We use the algorithm Adam with a learning rate of 0.001 and a batch size of 256 to minimize the objective function defined by equation \ref{eq:ELBO_prodLDA}. The training lasts at most 100 epochs, and is stopped if overfitting is detected in the evolution of validation loss.


\label{sec:training}

\subsection{Evaluation}

\begin{figure*}[!ht]
\RawFloats
\centering
\begin{minipage}{.30\linewidth}
    \centering
    \includegraphics[width=0.99\columnwidth]{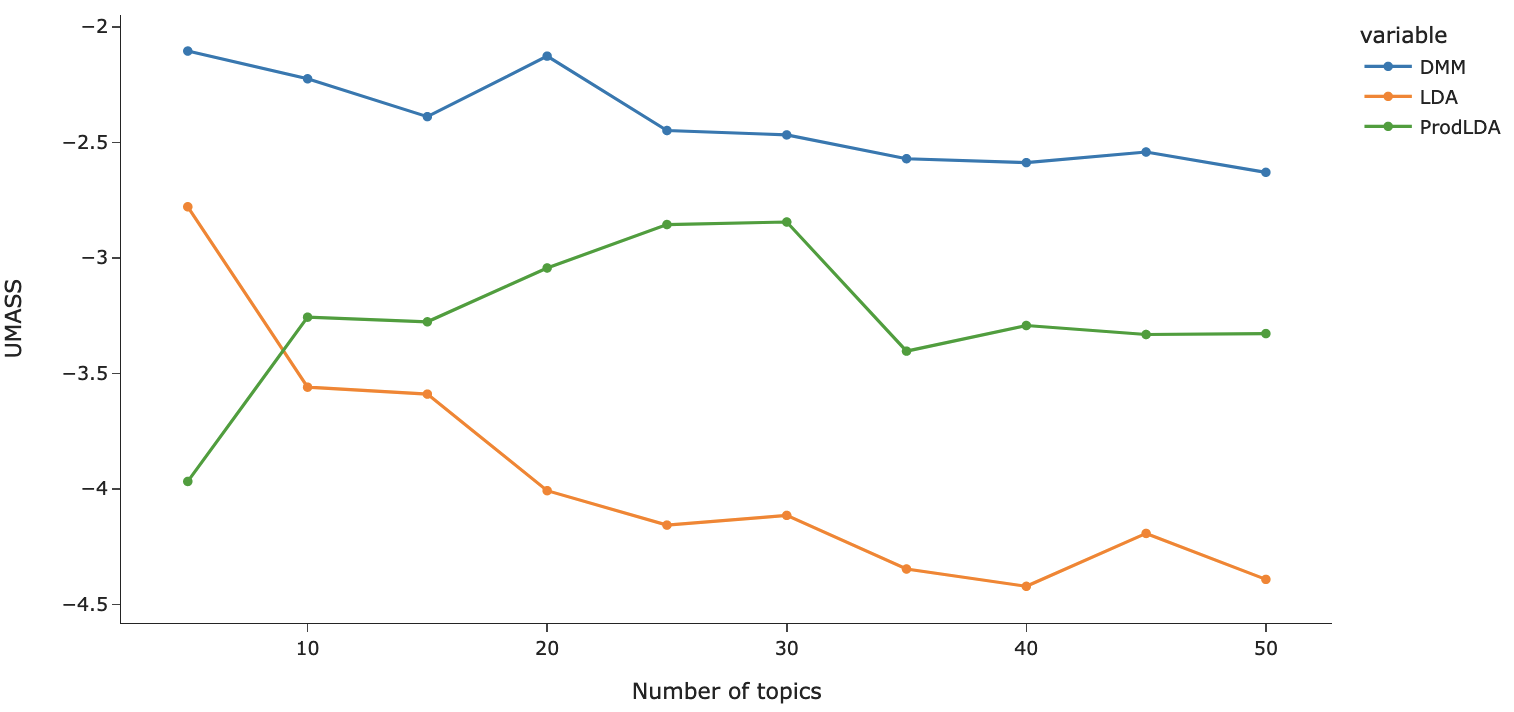}
    \caption{$U_{MASS}$ coherence on dataset \textbf{Bitcoin Talk}.}
    \label{fig:coherence_UMass}
    \end{minipage}%
    \hspace*{0.1cm}%
    \begin{minipage}{.30\linewidth}
    \centering
    \includegraphics[width=0.99\columnwidth]{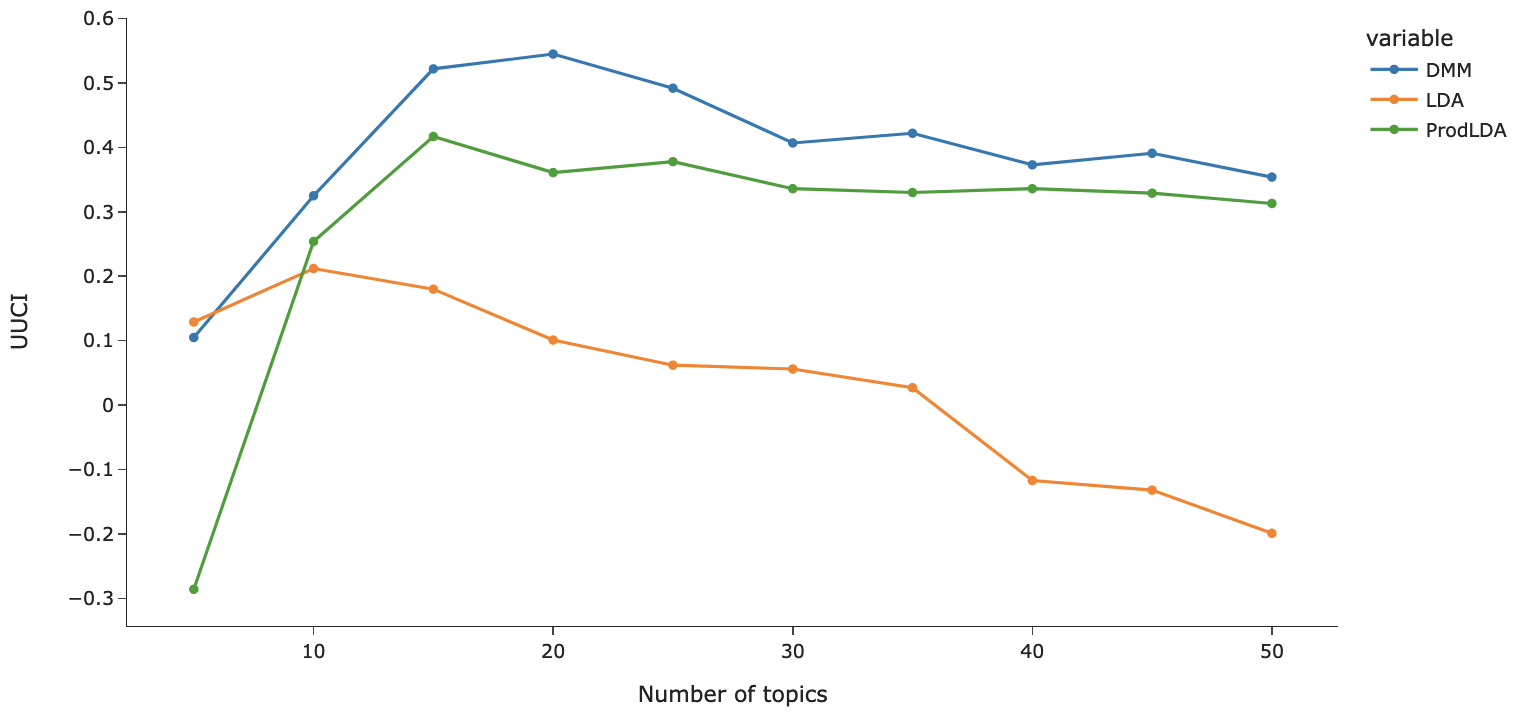}
    \caption{$U_{UCI}$ coherence on dataset \textbf{Bitcoin Talk}.}
    \label{fig:coherence_UCI}
    \end{minipage}%
    \hspace*{0.1cm}%
    \begin{minipage}{.30\linewidth}
    \centering
    \includegraphics[width=0.99\columnwidth]{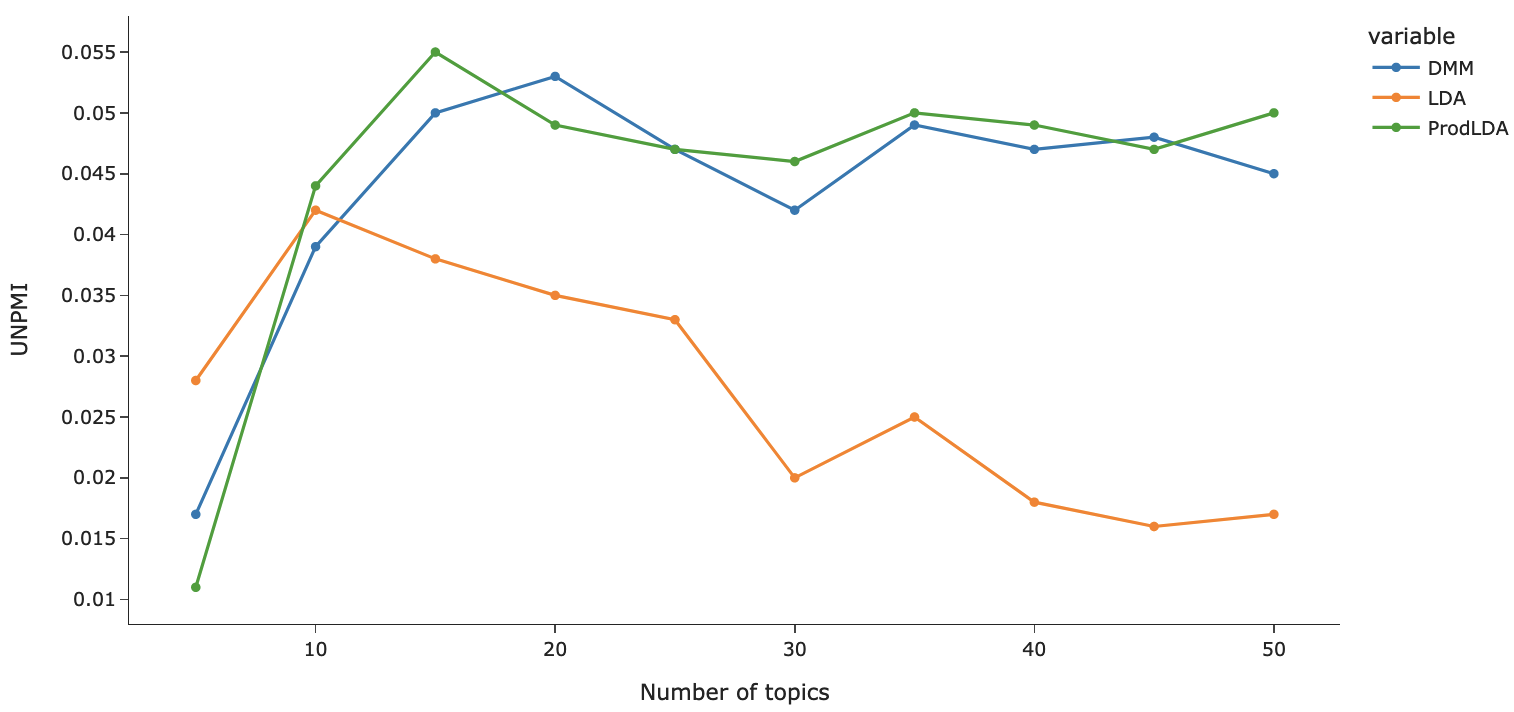}
    \caption{$U_{NPMI}$ coherence on dataset \textbf{Bitcoin Talk}.}
    \label{fig:coherence_UNPMI}
    \end{minipage}
\end{figure*}

As stated in the introduction, documents in the corpus \textbf{Bitcoin Talk} are globally longer that those in the dataset \textbf{Twitter}. For this reason, this corpus is more suitable for evaluating the different models. We use the coherence scores $U_{MASS}$, $U_{UCI}$ and $U_{NPMI}$ presented in section \ref{sec:eval}. The scores are calculated with $N=20$ and $w=20$ for $U_{UCI}$ and $U_{NPMI}$. Scores are reported in tables \ref{tab:coherence_UMass}, \ref{tab:coherence_UCL} and \ref{tab:coherence_UNPMI}, and in figures \ref{fig:coherence_UMass}, \ref{fig:coherence_UCI} and \ref{fig:coherence_UNPMI}.

\begin{table*}[!ht]
    \RawFloats
    \begin{minipage}{.33\linewidth}
      \centering
      \footnotesize
    \begin{tabular}{c|ccc}
     K    & DMM & LDA & prodLDA \\
         \hline
     5  & -2.105 & -2.779 & -3.968   \\
     10 & -2.225 & -3.560 & -3.257  \\
     15 & -2.389 & -3.590 & -3.277 \\
     20 & -2.127 & -4.008 & -3.044 \\
     25 & -2.449 & -4.157 & -2.856 \\
     30 & -2.468 & -4.115 & -2.845\\
     35 & -2.571 & -4.347 & -3.404 \\
     40 & -2.588 & -4.422 & -3.293\\
     45 & -2.542 & -4.193 & -3.332 \\
     50 & -2.630 & -4.392 & -3.328\\
     \hline
     Mean & $\mathbf{-2.408}$ & -3.956 & -3.260 \\
    \end{tabular}
    \caption{$U_{MASS}$ coherence \\ on dataset \textbf{Bitcoin Talk}.}
    \label{tab:coherence_UMass}
    \end{minipage}%
    \begin{minipage}{.33\linewidth}
     \centering
     \footnotesize
    \begin{tabular}{c|ccc}
     K    & DMM & LDA & prodLDA \\
         \hline
     5  & 0.105 & 0.129 &  -0.286 \\
     10 & 0.325 & 0.212 & 0.254 \\
     15 & 0.522 & 0.180 & 0.417\\
     20 & 0.545 & 0.101 & 0.361 \\
     25 & 0.492 & 0.062 & 0.378\\
     30 & 0.407 & 0.056 & 0.336  \\
     35 & 0.422 & 0.027 & 0.330 \\
     40 & 0.373 & -0.117 & 0.336\\
     45 & 0.391 & -0.132 & 0.329  \\
     50 & 0.354 & -0.199 & 0.313\\
     \hline 
     Mean & $\mathbf{0.394}$ & 0.032 & 0.277 \\
    \end{tabular}
    \caption{$U_{UCI}$ coherence \\ on dataset \textbf{Bitcoin Talk}. }
    \label{tab:coherence_UCL}
    \end{minipage}%
    \begin{minipage}{.33\linewidth}
     \centering
     \footnotesize
    \begin{tabular}{c|ccc}
    K    & DMM & LDA & prodLDA \\
         \hline
     5  & 0.017 & 0.028 & 0.011 \\
     10 & 0.039 & 0.042 & 0.044 \\
     15 & 0.050 & 0.038 & 0.055 \\
     20 & 0.053 & 0.035 & 0.049\\
     25 & 0.047 & 0.033 & 0.047 \\
     30 & 0.042 & 0.020 & 0.046 \\
     35 & 0.049 & 0.025 & 0.050 \\
     40 & 0.047 & 0.018 & 0.049\\
     45 & 0.048 & 0.016 & 0.047 \\
     50 & 0.045 & 0.017 & 0.050 \\
     \hline 
     Mean & 0.044 & 0.027 & $\mathbf{0.045}$ \\
    \end{tabular}
    \caption{$U_{NPMI}$ coherence \\ on dataset \textbf{Bitcoin Talk}.}
    \label{tab:coherence_UNPMI}
    \end{minipage} 
\end{table*}

In table \ref{tab:top_relevance}, we report for the model DMM trained on $K=20$ topics the top words for each cluster according to the relevance score introduced in section \ref{sec:human_judgement} with $\lambda$ equal to 0.3.

\begin{table*}
\footnotesize
    \centering
    \begin{tabular}{c | c}
         Topic & Top words \\
         \hline
         0 & maximalist, people, bitcoin, understand, scam, toxic, bitcoiner, think, shitcoin, criticism \\
         1 & friend, family, guy, dad, dinner, tell, buy, wife, pizza, man, girl \\
         2 & dump, panic, pump, sell, buy, fomo, crash, market, whale, btc \\
         3 & system, freedom, government, monetary, central, money, society, human, opt, nation \\
         4 & alt, profit, btc, trade, position, long, altcoin, pair, leverage, loss \\
         5 & rsi, resistance, daily, close, div, candle, hourly, bearish, bullish, bounce \\
         6 & value, layer, sov, store, settlement, censorship, utility, programmable, resistant, moe \\
         7 & s2f, asset, correlation, gold, sandp, stock, model, correlate, return, flow \\
         8 & episode, podcast, host, magazine, discuss, join, speaker, guest, topic, live \\
         9 & withdraw, pay, card, merchant, convert, ira, load, bitpay, payment, taxis \\
         10 & debt, print, fed, inflation, purchasing, rate, trillion, central, estate, bond \\
         11 & wallet, node, lightning, coinjoin, electrum, privacy, tor, hardware, wasabi, server \\
         12 & fork, consensus, segwit, segwit2x, contentious, uasf, reorg, core, pow, block \\
         13 & tender, legal, salvador, court, ban, india, wright, china, police, orea \\
         14 & upwards, breakout, resistance, area, zone, retest, diagonal, continuation, triangle, btcusd \\
         15 & cycle, precede, bull, correction, halve, retrace, prior, halving, historically, ath \\
         16 & inflow, outflow, volume, bitmex, gbtc, premium, exchange, greyscale, whale, bitfinex \\
         17 & renewable, mining, energy, electricity, miner, difficulty, hydro, hash, solar, subsidy \\\
         18 & tudor, stanley, sachs, goldman, manager, jones, paul, blackrock, fund, morgan \\
         19 & musk, microstrategy, tesla, elon, sheet, dorsey, balance, square, saylor, jack 
    \end{tabular}
\caption{Model: DMM. $K=20$ topics. Top words according to the relevance score of equation \ref{eq:relevance} with $\lambda=0.3$.}
\label{tab:top_relevance}
\end{table*}

\subsection{Application}
In this section, we study the predictability of bitcoin returns with data from Twitter. In order to use traditional machine learning algorithms, we first have to find an Euclidean representation of each tweet.  We set the number of topics $K$ equal to 20 and train the DMM with the optimal hyper-parameters $\alpha$ and $\beta$ found in section \ref{sec:training}. We recall that a tweet is assigned to only one topic in this model. For this reason, one-hot encoding is a natural choice to construct a tweet representation $t_{\mathrm{tweet}} \in \{0, 1\}^K$ as follows: for all $i \in \{0, 1, 2, ..., K-1\}$, $t_{\mathrm{tweet}, i} = 1$ if $i$ is the topic label of the tweet and 0 otherwise. In several previous studies, it has been demonstrated that sentiment of tweets could also give information about future movements of bitcoin market. To complete this application, we extend the representation $t_{\mathrm{tweet}}$ with two sentiment scores: a positive sentiment $p_{\mathrm{tweet}} \in [0, 1]$ and a negative sentiment $n_{\mathrm{tweet}} \in [0, 1]$, both are computed with the Python package \textsc{Vader} (\url{https://github.com/cjhutto/vaderSentiment}). $p_{tweet}$ and $n_{tweet}$ are ratios for proportions of text that fall in positive and negative vocabulary, respectively. In this application, we seek to predict the next daily return $r_{d+1}$ of bitcoin relatively to the token USDT from the set of tweets $\mathcal{T}_{d}$ published on day $d$. Quotes have been downloaded from the Binance API. For each day $d$, we compute a topic vector $t_{d}$ by averaging topic vectors over all tweets in $\mathcal{T}_d$ (equation \ref{eq:avg_top_vector}), we do the same to compute two average sentiment scores $p_d$ and $n_d$ (equations \ref{eq:avg_pos} and \ref{eq:avg_neg}).

\begin{equation}
\small
    t_{d} = \frac{ \sum_{\mathrm{tweet} \in \mathcal{T}_d} t_{\mathrm{tweet}}}{| \mathcal{T}_d |}
    \label{eq:avg_top_vector}
\end{equation}

\begin{equation}
\small
    p_{d} = \frac{ \sum_{\mathrm{tweet} \in \mathcal{T}_d} p_{\mathrm{tweet}}}{| \mathcal{T}_d |}
    \label{eq:avg_pos}
\end{equation}

\begin{equation}
\small
    n_{d} = \frac{ \sum_{\mathrm{tweet} \in \mathcal{T}_d} n_{\mathrm{tweet}}}{| \mathcal{T}_d |}
    \label{eq:avg_neg}
\end{equation}

$t_d$ is thus the empirical distribution of inferred subjects of tweets published within the day $d$. $p_d$ and $n_d$ are the average ratios of positivity and negativity in the tweets.  We train a linear regression model and report the results of the regression in table \ref{tab:reg_results_24}, in particular the p-values of significance t-tests for the different variables involved in the model. 

 \begin{table}
    \centering
    \tiny
    \begin{tabular}{c |c c c c c}
    Name    & variable & coef. & std. err. & t-stat & p-value  \\
    \hline
    Positive sentiment $**$ & $p_d$ & $0.200$ & $0.041$ & $4.819$ & $0.000$ \\
    Negative sentiment $**$ & $n_d$ & -0.249 & 0.047 & -5.317 & 0.000 \\
    \hline 
    Topic 0 ${**}$ & $t_{d, 0}$ & 0.040 & 0.009 & 4.263 & 0.000 \\
    Topic 1 ${**}$ & $t_{d, 1}$ & 0.097 & 0.012 & 7.779 & 0.000 \\
    Topic 2 ${**}$ & $t_{d, 2}$ & 0.046 & 0.014 & 3.407 & 0.001 \\
    Topic 3 ${**}$ & $t_{d, 3}$ & -0.066 & 0.018 & -3.611 & 0.000 \\
    Topic 4 ${**}$ & $t_{d, 4}$ & 0.038 & 0.011 & 3.311 & 0.001 \\
    Topic 5 ${**}$ & $t_{d, 5}$ & 0.037 & 0.011 & 3.179 & 0.001 \\
    Topic 6 ${**}$ & $t_{d, 6}$ & -0.046 & 0.015 & -3.159 & 0.002 \\
    Topic 7 ${**}$ & $t_{d, 7}$ & -0.048 & 0.010 & -4.897 & 0.000 \\
    Topic 8 ${**}$ & $t_{d, 8}$ & -0.118 & 0.014 & -8.693 & 0.000 \\
    Topic 9 ${**}$ & $t_{d, 9}$  & -0.202 & 0.018 & -11.236 & 0.000 \\
    Topic 10 ${**}$ & $t_{d, 10}$  & 0.030 & 0.011 & 2.661 & 0.008 \\
    Topic 11 & $t_{d, 11}$  & 0.021 & 0.016 & 1.347 & 0.178 \\
    Topic 12 ${**}$ & $t_{d, 12}$  & -0.032 & 0.012 & -2.725 & 0.006 \\
    Topic 13 & $t_{d, 13}$  & 0.026 & 0.014 & 1.914 & 0.056 \\
    Topic 14 ${*}$ & $t_{d, 14}$  & 0.055 & 0.027 & 2.029 & 0.042 \\
    Topic 15 & $t_{d, 15}$  & 0.016 & 0.014 & 1.100 & 0.271 \\
    Topic 16 & $t_{d, 16}$  & 0.010 & 0.012 & 0.800 & 0.424 \\
    Topic 17 & $t_{d, 17}$  & 0.014 & 0.015 & 0.930 & 0.352 \\
    Topic 18 & $t_{d, 18}$  & 0.022 & 0.016 & 1.404 & 0.160 \\
    Topic 19 & $t_{d, 19}$  & 0.029 & 0.013 & 2.145 & 0.032 \\
    \end{tabular}
    \caption{Regression results. \textbf{coef} : value of the coefficient in the model,  \textbf{std. err.} : estimation of the standard deviation of \textbf{coef}, \textbf{t-stat} : t-statistic, \textbf{p-value} : p-value of the t-statistic, $*$ : significant, $**$ : highly significant.}
    \label{tab:reg_results_24}
\end{table}

\section{Discussion}

It appears from the tables \ref{tab:coherence_UMass}, \ref{tab:coherence_UCL} and \ref{tab:coherence_UNPMI} that the model DMM globally  gives the most semantically coherent topics according to the three coherence scores evaluated on the dataset \textbf{Bitcoin Talk}. It is followed by the model prodLDA and finally by the model LDA, which is quite disappointing. It underlines that it is particularly difficult for the model LDA to deal with short texts. These results support the findings of \cite{yin2014dirichlet} and \cite{srivastava2017autoencoding}, in which  DMM and prodLDA were respectively introduced. We recall that documents are supposed to be generated from an unique topic in the model DMM. This assumption may be more reasonable for short texts compared to the assumptions of LDA and prodLDA. \\

According to tables \ref{tab:coherence_UMass}, \ref{tab:coherence_UCL} and \ref{tab:coherence_UNPMI}, setting $K$ equal to 20 is a good choice for training DMM. Top words according to relevance $r_{0.3}$ are reported in table \ref{tab:top_relevance}. Topics seem very semantically coherent: topics 5 and 15 are about technical analysis, topic 10 about monetary policies, topic 13 about regulation, topic 17 about mining and energy, topic 18 about actors of the Traditional Finance. In addition, topics are quite distinct from each other and cover a wide range of themes. \\

Concerning our application, we point out that several "topic variables", as well as sentiment variables, are highly significant for the prediction of a next day. This result demonstrates that topics discussed on Twitter could be informative about future price movements of bitcoin. Surprisingly, it appears that some topics that seem in favor of bitcoin, e.g. topics 3, 6, or 7, have negative coefficient, one hypothesis is that market pauses for breath after this kind of positive news for Bitcoin. Finally, a negative sentiment globally impacts negatively the next day return, and, inversely, a positive sentiment impacts positively the next day return. \\

\section{Conclusion}
Of all the models that have been trained on the corpus of tweets about bitcoin, the \textit{Dirichlet Multinomial Mixture} (DMM) is the best-performing model according to three coherence scores for most choices of $K$, the number of topics. Setting $K$ equal to 20 allows to extract semantically coherent and non-redundant topics. In addition, these topics are easily understandable for human experts. Finally, representations built from a trained topic model can be used for real-world applications, such as predicting future movements of bitcoin.

\bibliographystyle{IEEEtranN}
\bibliography{bib}

\end{document}